# X-rays from HH 80, HH 81, and the Central Region

Steven H. Pravdo
Jet Propulsion Laboratory, California Institute of Technology
306-431, 4800 Oak Grove Drive, Pasadena, CA 91109; spravdo@jpl.nasa.gov

Yohko Tsuboi
Department of Physics, Faculty of Science and Engineering, Chuo University,
Kasuga 1-13-27, Bunkyo-ku, Tokyo 112-8551, Japan; tsuboi@phys.chuo-u.ac.jp

and

Yoshitomo Maeda
Department of High Energy Astrophysics, Institute of Space and Astronautical Science, Japan Aerospace Exploration Agency, 3-1-1 Yoshinodai, Sagamihara, Kanagawa 229-8510, Japan; ymaeda@astro.isas.jaxa.jp

## ABSTRACT

We report detections of X-rays from HH 80 and HH 81 with the ACIS instrument on the *Chandra* X-ray Observatory. These are among the most luminous HH sources in the optical and they are now the most luminous known in X-rays. These X-rays arise from the strong shocks that occur when the southern extension of this bipolar outflow slams into the ambient material. There is a one-to-one correspondence between regions of high X-ray emission and high Hα emission. The X-ray luminosities of HH 80 and HH 81 are 4.5 and 4.3 x $10^{31}$ erg $s^{-1}$, respectively, assuming the measured low-energy absorption is not in the sources. The measured temperature of the HH plasma is not as large as that expected from the maximum velocities seen in the extended tails of the optical emission lines. Rather it is consistent with the ~$10^6$ K temperature of the “narrow” core of the optical lines. There is no observed emission from HH 80 North, the northern extension of the bipolar flow, based upon a measurement of lower sensitivity. We imaged the central region of the bipolar flow revealing a complex of X-ray sources including one near, but not coincident with the putative power source in the radio and infrared. This source, CXOPTM J181912.4-204733, has no counterparts at other wavelengths and is consistent in luminosity and spectrum with a massive star with $A_V$ ~ 90 mag. It may contribute significantly to the power input to the complex. Alternatively, this emission might be extended X-rays from outflows close to the power source. We detect 94 X-ray sources overall in this area of star formation.

## 1. INTRODUCTION

Herbig-Haro objects are a new class of X-ray sources. Recently observers detected X-rays from two HH objects associated with protostellar outflows. These are HH 2 (Pravdo et al. 2001) and HH 154 (Lynds 1551 IRS5, Favata et al. 2002). The positions of the X-rays tell us where the shocks are strongest. The spectra of the X-rays indicate the temperatures and the elemental abundances in the shocked gas. The X-ray luminosities and morphologies of the emitting regions reveal the densities of the X-ray emitting gas.

HH objects are optically luminous knots of material first discovered ~50 years ago in star-forming regions (Herbig 1951, Haro 1952, Reipurth & Bertout 1997). An identifying characteristic is their high proper motions often oppositely directed from a central point of origin. It is believed that young central stars or proto-stars power the outflows via jets. HH objects form at the shock fronts when jet material smashes into the interstellar medium. Multi-wavelength studies conclude that HH objects represent regions wherein shock-heated material cools. The characteristic emissions in these recombination zones include Balmer lines, [SII], and [OIII] and indicate temperatures of $10^4$-$10^5$ K (e.g. Böhm 1956). In the case of a few HH objects with high excitation and low extinction such as HH 1/2, radio free-free emission is also detected (Pravdo et al 1985). A missing ingredient in this picture had been the detection of the T ~ $10^6$ K X-ray emission expected from the high velocities of some outflows.

In HH 2 the observed T ~ $10^6$ K X-rays arise from a shock front that also exhibits strong H$\alpha$ and radio continuum emission, and that is well downstream from the origin of the outflow (Pravdo et al. 2001). The X-rays from HH 154 are consistent with emission from a shock at the base of its protostellar jet, but might have other origins (Bally et al. 2003).

Reipurth & Graham (1988) discovered HH 80 and 81. These objects are in the "HH 80/81 complex," an extended region near RA = $18^h$ $19.1^m$, Dec.= -20° 52'. Martí, Rodríguez, & Reipurth (1993, hereafter MRR93) propose a unifying structure for the complex based upon their VLA observations. They suggest that the central power source is IRAS 18162-2048, estimated it to be a pre-main sequence (PMS) star with a luminosity of ~$10^4$ $L_\odot$, and that it is associated with their VLA source 14 (hereafter MRR 14). In the words of MRR93: "The morphological characteristics of the HH 80-81 jet are very similar to those of jets found in association with low-mass stars, except for being an order of magnitude larger. This result suggests that the mechanism which produces highly collimated jets in young low-mass stars also operates in new born massive stars, thousands of times more luminous." The quasi-linear outflows extend 5.0' to the southwest (HH 80 and 81) and 6.2' to the northeast (HH 80 North), or 2.5 pc and 3.1 pc, respectively, at the nominal distance of 1.7 Kpc (MRR93). These extents are >10 times larger than those of other known outflows, and are consistent with the higher luminosities in the optical emission (Heathcote, Reipurth, & Raga 1998, hereafter HRR98).

The HH 80/81 complex is located within the giant molecular cloud discovered by Saito et al. (1999) that is a site of active star formation. The cloud also contains IRAS 18162-2048 = MRR 14 = GGD 27-IRS (see §4.3). GGD 27-IRS has been extensively studied in the infrared. Aspin et al. (1991) and Aspin & Geballe (1992) propose that it is a star formation region and that several intermediate- to high-mass young stellar objects are responsible for its heating.

We present the results of a *Chandra* observation of the HH 80/81 complex. The observation is focused on the bright HH objects 80 and 81 in the southern extension of the complex, but it also includes observations of the northern extension, and of the central region believed to contain the power source. We use these to further understand interactions of protostellar outflows with their ambient medium, the underlying cause of the HH phenomena, and to discover more about this complex galactic region.

## 2. THE OBSERVATIONS AND ANALYSIS

We searched for X-ray emission from the HH 80/81 complex in a 37.3-ks *Chandra* X-ray Observatory (Weisskopf et al. 2002) observation on November 4, 2002. Because we expected the X-ray emission to be predominantly < 1 keV, we used the ACIS-S configuration with the back-side illuminated chip S3 on the X-ray "optical" axis. ACIS-S was in the "faint" data mode with 3.2-s frame readouts. We examined data from HH 80/81 on two chips: S3 and S4 using the level 2 processed data (or equivalent on S3) with the standard definitions for good events. For S3 we made charge-transfer-inefficiency (CTI) corrections supplied by a procedure developed at Pennsylvania Sate University (Townsley et al. 2000, Townsley 2003). S4 is a front-side illuminated chip and the results of the standard processing were used after "destreaking" as described in *Chandra* data analysis system, CIAO Science Threads (CIAO 2003). Figure 1 shows the filtered image of S3 and S4. We show the locations of the prominent celestial regions and some sources discussed in the text.

We performed positional, timing, and spectral analyses of these data. We used a wavelet-based source detection program (Freeman et al. 2002) to locate X-ray sources on S3 in a soft band, 0.2-2.0 keV, and in a hard band, 2.0-8 keV, and on S4 in the total energy band, 0.4-8 keV. We performed encircled energy analysis by examining the spatial distribution of detected sources around their centroids. We constructed light curves to check for source variability.

Finally, we used XSPEC to measure the spectra of the detected sources. Source counts were extracted from regions around sources that matched the local point-spread function (PSF), and backgrounds were taken from nearby source-free regions. The instrumental response matrices are deemed to be reliable > 0.4 keV, so we only examined spectra in the range 0.4-8 keV. Above 8 keV there were no significant counts. The spectral models included the functions "WABS" for interstellar absorption, and either "MEKA-L" for a thermal spectrum (Mewe, Gronenschild, & van den Oord 1985) or a powerlaw for the continuum. To account for the observed quantum efficiency decay of ACIS, possibly caused by molecular contamination of the ACIS filters, we applied a time-dependent correction to the ACIS quantum efficiency implemented in XSPEC with the model "ACISABS," a model contributed to the *Chandra* users software exchange worldwide web site (Chartas 2003). Fixed default parameters were used for ACISABS except that the time from launch was set at 1200 days, appropriate for our observation.

We calculated the 1-σ limits for spectral parameters by first allowing all the parameters to vary simultaneously to find the best-fit values. Then a parameter of interest was fixed and we stepped through different values of that parameter, allowing all the other parameters to vary. The value of the interesting parameter when chi squared increased by 3.5 is the 1-σ limit for a fit with 3 free parameters (Lampton, Margon, & Bowyer 1975).

We checked the astrometric accuracy of the data by comparing the positional output of the source detection program with the positions of unambiguously identified stellar counterparts found in the USNO B1 catalog. On S3 there were 34 stars located with 1.5” of soft X-ray sources and 13 stars located within 1.5” of hard X-ray sources. On S4 only 3 stars are found within 1.5” of X-ray sources. For the soft S3 sources the average X-ray-optical offsets are 0.08±0.67” in RA and 0.11±0.59” in Dec. For the hard S3 sources, the offsets are 0.21±0.45” in RA and -0.16±0.67” in Dec. In S4, the offsets are 0.24±0.53” in RA and -0.24±0.20” in Dec. There are no systematic offsets in the X-ray astrometry for the on-axis S3 and not enough evidence to make any corrections for S4.

## 3. RESULTS

### 3.1 X-ray Sources

Table 1 gives a list of the 94 X-ray sources in this region detected with at least 2-σ significance. The Chandra (CXOPTM) sources were compared with the USNO B1 Catalog (Monet et al. 2003) and the 2MASS Catalog. The majority of the sources are believed to be associated with stars and stellar phenomena found in star formation regions (SFRs, e.g. Ku & Chanan 1979, Feigelson & DeCampli 1981). The 76 S3 sources consist of 39 sources detected in the soft band only, 18 detected in the hard band only, and 19 detected in both. Six of the latter have higher count rates in the hard than soft band. It is likely that the apparent spectral differences are in many cases not intrinsic, but due to their positions in the SFR—those sources on the near side of the molecular cloud (Yamashita et al. 1989) are less absorbed and appear softer. An indication of this is that 21 of the 24 sources that dominate in the hard band also are in blank optical/IR fields. However, sources associated with HH objects (see below) are intrinsically softer than stellar sources since they are not significantly absorbed, yet have weak or no hard components. The S3 count rates are given in separate soft (“s,” 0.2-2 keV) and hard (“h,” 2-8 keV) bands in Table 1, while the S4 count rates are given in full band (but listed in the “s” column with an “f” in the “h” column, 0.4-8 keV). Typical conversions from counts to intensity (units are ergs $cm^{-2}$ $sec^{-1}$ per counts $ksec^{-1}$) in the three bands are: 3.1 x $10^{-15}$, 0.47-3.5 x $10^{-14}$, and 2.0 x $10^{-14}$, for “s,” “h,” and “f,” respectively. The large range in the conversion values for the “h” band is exemplified by bright, soft sources such as SS 365 at the low end, and hard sources such as CXOPTM J181905.4-205202 (see Table 2 and below) at the high end. Upper limits for non-detections are 1.6 x $10^{-16}$, (0.28-2.1) x $10^{-15}$, and 1.7 x $10^{-15}$ ergs $cm^{-2}$ $sec^{-1}$ for “s,” “h”, and “f,” respectively.

### 3.2 HH 80 and 81

The optically brightest HH knots produce extended X-ray emission. The two brightest knots, HH 80A and HH 81A, are the strongest X-ray sources. Figures 2 and 3 show the portions of the ACIS-S3 images centered on HH 80 and HH 81, respectively, superimposed upon Hα profiles (contours) from Hubble Space Telescope (HST) archived data (WFPC2 Associations 2003). We also show the VLA sources from MRR93. HH 80A and HH 81A are detected with 46 counts and 63 counts, respectively. In addition we detect the X-rays from the smaller knots, HH 80G,H with 17 counts, and HH 80 C,D with 6 counts. The ACIS sources in these fields are indicated in the figures and listed in Table 1. We constructed encircled energy versus radius plots for HH 81A, HH 80A, and HH 80G/H to measure the spatial distributions of the X-rays (Figure 4). We compared these to two suspected point sources nearby, CXOPTM J181901.8-205242 = SS 365 and a new X-ray source, CXOPTM J181905.4-205202 (HARD), in Figure 2. SS 365 is a V=12.5 weak-Hα emission-line star, estimated to have spectral type A0 (Stephenson & Sanduleak 1977). The spatial profiles of these two are statistically alike and indicative of the near on-axis ACIS/*Chandra* PSF. The HH objects have distinguishably flatter spatial profiles than either of the two comparison sources, indicating that the HH X-ray emission is extended. The HH 80A and HH 81A distributions differ from the point sources at > 99% confidence using the K-S test. The lower counts in HH 80 G/H reduce this confidence to 97%.

We constructed light curves with the data in 5000-s bins. There were no indications of variability from any of the knots.

We examined each of the spectra from the HH 80 and 81 knots and found them to be consistent with each other. Therefore to achieve the highest statistical accuracy we combined the data from all the knots to construct a composite HH 80/81 spectrum. The source region used for this composite spectrum was larger than that used for the individual detected sources (and included counts not detected as sources) resulting in higher net counts. We fit this spectrum using a MEKA-L model with instrumental and interstellar absorption. The fit was acceptable with $X^2$=22 for 22 degrees of freedom (Figure 5). The best-fit plasma temperature is kT = 0.13 ± 0.05 keV with a neutral hydrogen column density of $N_H$ = 4.4 (+ 1.0,-1.2) x $10^{21}$ $cm^{-2}$. The elemental abundance in the plasma has a 1-σ lower limit 0.04 of the standard cosmic abundance, while the upper limit is not interestingly constrained. The observed intensity between 0.4-2.0 keV is 1.2 x $10^{-14}$ ergs $cm^{-2}$ $s^{-1}$. Table 2 lists the spectral results.

The Hα contours shown in Figures 2 and 3 were obtained with HST about 7 years before the Chandra observations. The proper motions of the brightest knots are expected to be ~0.07" and 0.25" southwest based upon the measurements of HRR98. This would result in a negligible adjustment of the overlays shown. The proper motion of these radio knots has not been measured but the knots in the central source are seen to move as much as 0.16" $y^{-1}$ (Martí, Rodríguez, and Reipurth 1995), so the ~12-y gap between the radio and X-ray observations could explain the ~1" offsets to the northeast of the sources MRR 7 and 8.

### 3.3 Hard Source near HH 80

CXOPTM J181905.4-205202 (HARD) is a newly discovered X-ray source with no counterpart at other wavelengths, shown in Fig. 2. It is distinguished from the HH X-rays sources nearby in two ways. First, as indicated in Figure 4, it appears to be a point source, similar to that seen in SS 365. Second, its spectrum is harder, fit equally well with a high temperature thermal spectrum or a powerlaw spectrum with photon index of ~1 (Figure 6, Table 2). There is no significant temporal variability in the 30 counts arising from this source.

### 3.4 The Central Region

The X-ray source near IRAS 18162-2048 = MRR 14, is situated on the edge of the S3 chip ~5' off-axis. The PSF is, as expected, considerably more extended and asymmetric than that of the on-axis sources and the positional accuracy is degraded to >1". Furthermore we cannot determine whether the X-ray source is resolved. Figure 7 shows the 2MASS $K_s$ image of GGD 27 that contains MRR 14 and a number of other well-studied radio and IR sources. We superimposed the ACIS sources with 0.1-, 0.2-, and 0.5-counts contour levels (background is 0.04 counts), the MRR93 VLA sources, and some of the IR sources described by Stecklum et al. (1997). The X-ray contours show a ~5" northwest-southeast elongation due to the shape of the off-axis PSF convolved with any intrinsic extent. CXOPTM J181912.4-204733 is centered about 4" southeast of the VLA/IR source position. This source falls upon a blank field in optical, IR, and radio. See Table 2 for the spectral parameters. There is no evidence for variability.

X-ray emission appears to extend, at a low level, to the western side of MRR 14. In this "western lobe" of the emission there are 2 known IR sources, IRS 8 and 9. Indeed, as discussed below, the source detection algorithm found a source at the position of IRS 8. The number of counts in this source is ~8. This is far below the ~35 counts that we find by summing the net counts in a 10" box centered on this lobe. A look at the spectrum of these counts shows that they are neither < 2 keV like the HH spectra, nor strongly absorbed like CXOPTM J181912.4-204733, to the east. A more sensitive observation is required both to determine whether there is extended emission here and to measure its spectrum.

The X-ray source, CXOPTM J181911.7-204727, is within 1" of the source HL 18 (Hartigan & Lada 1985, hereafter HL85) = IRS 8 (Stecklum et al. 1997), and may be identified with it. The source CXOPTM J181912.0-204703 is located in the west wing of the "wishbone" in Figure 7, without any counterparts at other wavelengths. It is detected in the hard band only.

Two other ACIS sources are coincident to <1" and <3" with the VLA positions of MRR 12 = HL 41 = IRS 4 (Yamashita et al. 1987) and MRR 32 = GGD 28B = HL 8, respectively. The first of these, CXOPTM J181910.4-204657, shown in Figure 7, is coincident with MRR 12. The second, CXOPTM J181921.8-204535, is located further east and north, on the edge of chip S4. Figure 8 shows the X-ray source and contours at 0.04, 0.3 counts (background is 0.01 counts), MRR 12, and HL 8. The centroid of the X-ray emission is offset from the radio position by ~2.7", and is ~1.7" from one of the two 2MASS sources that comprise the $K_s$ emission from this location. However, again, this object is further off-axis than the objects on CCD 7. Therefore the PSF for this emission is even more extended, as shown by the contours, and the positional resolution is worse. The source is likely to be associated with the radio position. The spectra of these X-ray sources are described in Table 2. Both fit thermal models that are hotter than the HH objects. Their light curves show no evidence for variability.

### 3.5 HH 80 North

HH 80 North was located on the S4 chip. This chip has far lower sensitivity to X-rays < 1 keV under the best circumstances, and in addition loses sensitivity due to the instrumental contamination that causes streaking. There was a slight excess of ACIS counts at the position of HH 80 North: about 1 count within 3" radius and 5 counts within a 5" radius compared with a large nearby source free-region. We estimate the upper limit to X-ray emission from HH 80 North to be, ~2 x $10^{-15}$ ergs $cm^{-2}$ $sec^{-1}$, by comparing this "excess" count with the count rates and fluxes of detected sources on S4. This is ~6 times smaller than the combined, observed intensity of the southern HH objects (see Table 2).

### 3.6 Other Sources

Figure 9 shows the JHK color-color diagram for the identified sources in this region. The interesting sources discussed in the text are labeled. SS 365 = CXOPTM J181901.8-205242 shows no evidence for variability and fits a thermal spectrum with parameters shown in Table 2. It and other reddened stars such as CXOPTM J181911.7-204727 (IRS8) and J181906.0-205115 (Hα near HH 81) appear in the same JHK color region as classical T Tauri stars (Lada & Adams 1992). This is discordant with its "weak Hα" classification. CXOPTM J181906.2-204232 is a bright discovery, within 0.4" of the USNO B1 star 0692-0595725 with *B*=19.3, *V*= 15.48 (USNO A2 has 18.3, 15.6). It is also 2MASS 18190626-2042320. This is the only source that shows strong evidence of temporal variability (Figure 10) with either a flare or a partial cycle of periodic behavior. Its IR colors show it to be quite red ("Variable" in Figure 9). Based upon its color and X-ray luminosity, 2.5 x $10^{31}$ ergs $s^{-1}$ assuming it is at 1.7 Kpc, this is a PMS star, e.g. a Herbig Ae/Be, rather than an X-ray binary.

The "bluest" source in the diagram is CXOPTM J181907.1-205349. It may be a foreground object as it shares the colors of a main sequence G star. If it were at the distance of HH 80/81 it would have an X-ray luminosity of ~2.5 x $10^{29}$ ergs $s^{-1}$.

## 4. Discussion

### 4.1 X-ray emission from HH Objects

With the discovery of X-ray emission from HH 80/81 it is now clear that the proto-stellar outflows are a new class of X-ray sources, at least, those identified with the most energetic HH objects. This modest-length (37.3 ks) *Chandra*/ACIS observation with its ~1" spatial resolution, reveals X-rays associated with all the brightest optical emission lines of the HH 80/81 complex (see Figures 2 & 3). A more sensitive observation, we predict, would show an extension of the X-ray emission over the entirety of the Hα regions.

The X-ray emission from HH 80/81 is larger than that predicted from a simple extrapolation of HH 2H. With the HH 80/81 luminosities in the optical and radio higher by ~10 and the distance to HH 80/81 larger by ~3.5, we expected roughly equal X-ray intensities. Instead, the unabsorbed intensity from HH 80/81 is ~2.6 x $10^{-13}$ ergs $cm^{-2}$ $s^{-1}$, or ~60 times larger than HH 2 (0.36-2 keV). This is not attributable to higher shock velocities of HH 80/81 pushing the X-rays to higher temperatures or luminosities, since the observed temperature in HH 80/81, T = (1.5 ± 0.6) x $10^6$ K, is consistent with that of HH 2H.

This last point is somewhat surprising. The temperature implies a shock velocity of $v_s \sim (T/15)^{1/2} =$ 320 km $s^{-1}$ (Raga, Noriega-Crespo, & Velázquez 2002, hereafter RNV02). This velocity is considerably lower than the proper motions measured in the radio along the jet, 600-1400 km $s^{-1}$ (Martí, Rodríguez, & Reipurth 1995), or of the bowshock velocities, 650±50 km $s^{-1}$ (HH 80A) and 700±50 km $s^{-1}$ (HH 81A) inferred from an examination of the optical emission line profiles. At ~arcsec resolution the radio continuum, optical emission lines, and X-rays are co-located. How do these different emission components fit together in the knots?

HRR98 use spatially- and spectrally-resolved HST observations of the optical line emissions' faint high-velocity tails to infer the larger shock velocities. From this, and the measured H$\alpha$ surface brightness they derive densities of ~400 $cm^{-3}$ and cooling distances larger than the size of the knots. However, they note that if the faint high-velocity tails are excluded, then the maximum velocities are ~250 km $s^{-1}$. Since the X-ray results show no evidence of significant gas with these higher shock velocities, it appears that the faint tail emission should be excluded in the first order inference of the physical properties. Using the X-ray determined velocity of 320 km $s^{-1}$, then the optical gas density is ~1000 $cm^{-3}$ and the cooling distance, ~4 x $10^{15}$ cm, is less than the size of the knots, making them fully radiative (e.g. Raga & Böhm 1986).

MRR93 question whether the radio emission can be optically thin based on their measured spectral index, $\nu = -0.3 \pm 0.1$, and suggest that a synchrotron component is present in addition to the $\nu = -0.1$ optically thin free-free emission (Rodríguez et al. 1993). (It is interesting to note that the brightest and only X-ray-detected knot in HH 2, HH 2H = VLA 10--Rodríguez et al. 2000--also has a slightly more negative index, $\nu = -0.2 \pm 0.1$, than the other nearby knots.) The optically thin model of Gavamian & Hartigan (1998) for free-free radiation from shocks with $v_s \sim 320$ km $s^{-1}$ can explain the observed radio emission from HH 80 and 81 with preshock densities $\leq 10^3$ $cm^{-3}$, where the viewing angle is not constrained by the spectral index. Thus the optical and radio emission can arise from shocks with similar densities.

If we apply the analysis of RNV02 to the X-ray emission, we find X-ray densities of ~40 $cm^{-3}$ and ~80 $cm^{-3}$ for HH 80 and 81, respectively. The radiative and non-radiative solutions give similar answers, as the cooling distances are now comparable to the knot sizes. The inferred X-ray densities for free-free emission are ~10 times less than the optical and radio, implying perhaps smaller X-ray fill factors. Tables 3 and 4 show a compilation of measured and derived physical parameters for HH 80/81.

RNV02 model the X-ray emission from HH 80/81 using physical parameters deduced from the optical observations. As shown above, these are not entirely consistent with the physical parameters derived from the X-ray measurements. They assume $N_H$ = 3.0 x $10^{21}$ $cm^{-2}$ based upon the HRR98 extinction measurement, $A_v$ = 2.33. If we use the empirical relationship (e.g. Gorenstein 1975) between column density and extinction, $A_v$ =4.5 x $10^{-22}$ $N_H$, we obtain a larger $N_H$ = 5.2 x $10^{21}$ $cm^{-2}$, consistent with the X-ray measurement from the average spectrum, 4.4 (+1.0,-1.2) x $10^{21}$ $cm^{-2}$ (Table 2). RNV02 present two estimates for the intensity of HH 81 based upon "consider[ing] the extinction at 1 keV [or] at 0.5 keV…," or two different levels of extinction. Their estimates are: 7.5 x $10^{-15}$ and 2.5 x $10^{-12}$ erg $cm^{-2}$ $s^{-1}$. Their lower value is in reasonable agreement with our measurement of 4.7 x $10^{-15}$ erg $cm^{-2}$ $s^{-1}$. If we assume that all the absorption is interstellar, then the flux at the source becomes 1.3 x $10^{-13}$ ergs $cm^{-2}$ $sec^{-1}$ (0.4-2 keV) for HH 80 and 1.2 x $10^{-13}$ ergs $cm^{-2}$ $sec^{-1}$ (0.4-2 keV) for HH 81. At a distance of 1.7 Kpc the (unabsorbed) fluxes from HH 80 and 81 are 4.5 x $10^{31}$ and 4.3 x $10^{31}$ ergs $s^{-1}$, respectively. Both of these objects are nearly 100 times more luminous than HH 2H (Pravdo et al. 2001) or L1551 (Favata et al. 2002) and confirm the scaled up energetics of HH 80/81 at all wavelengths.

Figure 11 shows a plot of X-ray vs. radio emission for the four known X-ray HH objects. We also show the function $L_X \sim L_R^{1.4}$, the best fit for these data. The fit has a spectral index error of $\pm$0.3 and so is marginally consistent with a linear relationship or an index of 1.24 that Güdel (2002) shows is appropriate for $L_X$ vs. $L_R$ of stars, wherein the energy arises from the magnetic fields.

### 4.2 HH Object Morphology

HH 80 and 81 are located 2.5 pc from the putative exciting source, IRAS 18162-2048. The radio emission knots form highly collimated linear structures ($\geq$20 ratio for the parallel compared with perpendicular dimension), from the center toward HH 80/81 and in the opposite direction toward HH 80 North. MRR93 also suggest that there is a slightly sinusoidal path due to precession of the ejection axis.

Why are HH 80/81 the only sites of intense radio/optical/X-ray emission along the southern path of the linear outflow? A variation of the shocked cloudlet model (Schwartz 1978) offers the best explanation. In this case a jet from the power source encounters the dense cloudlets. The shocked jet material provides the high temperature X-ray emission while the heated cloudlets provides the optical emission.

An interesting detail of this flow is in the HH 80 complex. An arc of H$\alpha$ emission starts at HH 80A and extends counterclockwise to HH 80J,K (HRR98's Figure 5). This structure may be explained with the discovery of the hard X-ray source, CXOPTM J181905.4-205202, located near the focus of this arc (Figure 2). We suggest that the X-ray source is a buried protostellar object, considerably less energetic than IRAS 18162-2048, but still powerful enough to push material outward creating a low-density bubble on the western side of HH 80. Its luminosity is ~1.2 x $10^{31}$ erg $s^{-1}$ at the distance of 1.7 Kpc (where the luminosity in insensitive in this case to either the thermal or powerlaw models with or without absorption). This is near the projected edge of a high-density molecular region (HRR98). This putative proto-star may define the edge near HH 80.

CXOPTM J181906.0-205115 is a soft X-ray source coincident with Hα emission (Fig. 3). It is most likely a PMS Hα-star based upon its positional coincidence with a USNOB1 and 2MASS source (Table 1, Figure 9 "Hα near HH81").

### 4.3 X-ray Sources Near the Central Power Source

The identity of the source that ejected the outflows and created the HH objects is of great interest. A convergence of results at different wavelengths now point toward an unseen stellar object in the infrared, GGD 27-ILL (Aspin et al. 1991). It illuminates IRS 2, once believed to be self-luminous (Yamashita et al. 1989), but now seen to shine with reflected light. High-resolution 8.5-μm observations by Stecklum et al. (1997) confirm that GGD 27-ILL is the illuminator of the northern reflection nebula as well—the large wishbone-shaped region in Figure 7. GGD 27-ILL is coincident to within ~0.1" with the peak of the resolved emission from MRR 14, and is consistent with IRAS 18162-2048 and the extended 2MASS source 18191205-2047319. Is it the power source for the HH objects?

The centroid of the nearby X-ray source is significantly offset, ~4", from GGD 27-ILL. The X-rays are unlikely to be related to the objects mentioned above unless they arise from an extended source that we only view through the transparent edge of the absorbing cloud. Figure 7 shows that CXOPTM J181912.4-204733 is located off the edge of the wishbone nebula. The X-ray spectrum is unusual compared with the other spectra examined herein—highly absorbed by X-ray standards, suggesting that the source is seen through a significant column density of ambient material, ~2 x $10^{23}$ H atoms $cm^{-2}$. This corresponds to $A_V$ ~ 90 mag. The underlying X-ray spectrum has a temperature of ~1 x $10^7$ K, typical of PMS stars (e.g. Feigelson et al. 2002), harder than that of HH objects, and softer than that of CXOPTM J181905.4-205202 (see above). The unabsorbed luminosity is ~ 2 x $10^{32}$ ergs $s^{-1}$, quite high for a PMS star. In the Orion Nebula Cluster, for example, only $\Theta^1$C, an O6 star at ~$10^{33}$ ergs $s^{-1}$ is more luminous. If it is an intermediate-mass star with $L_x/L_{bol}$ typically < 2 x $10^{-4}$, it would have $L_{bol} \geq 2.5 \times 10^3\ L_\odot$. A high-mass star with $L_x/L_{bol}$ typically ~ $10^{-6}$ (Feigelson et al. 2002) would have $L_{bol} \sim 5 \times 10^5\ L_\odot$. The X-ray source may be revealing a massive star that contributes significantly to the energetics of the region. In keeping with this stellar source explanation, we would also explain the weaker emission from the "western lobe" as arising from individual PMS/proto-stars such as IRS 8.

An alternate explanation for the X-ray emission near MRR 14 is that we are seeing X-rays from a shock near the outflow source, similar to a possibility presented by Bally et al. (2003). They suggest that X-rays coincide with the base of the HH 154 jet. The central source in MRR 14 is known to contain fast moving jets (Marti, Rodríguez, & Reipurth 1998). We may view their X-rays preferentially in the eastern lobe where the absorption is relatively lower.

Two other noteworthy sources are CXOPTM J181911.7-204727 = HL 18 = IRS 8, and CXOPTM J181912.0-204703 that appears in line with the western branch of the wishbone nebula. Stecklum et al. (1997) describe IRS 8 as either a B-type star with $A_V$~9 mag or a cooler, foreground star. The soft band X-ray detection favors the later choice while the fact that IRS 8 is the extreme red object in Figure 9 favors the former. The other X-ray source is a hard source but there are not enough counts in the spectrum to determine whether it is a highly absorbed spectrum or an intrinsically hard spectrum.

### 4.4 X-ray Sources Associated with Two Other VLA Sources

GGD 27 and GGD 28 were originally suggested as HH objects (Gyulgudaghian, Glushkov, & Denisyuk 1978) but are not, as they have no nebular Hα emission (HL85). CXOPTM J181910.4-204657 is coincident (<1") with the GGD 27 object HL 41 = MRR 12 = IRS 4. It is not an Hα source (HL85), and is self-luminous (Yamashita et al 1987). The position of CXOPTM J181921.8-204535 is consistent with that of HL 8 = GGD 28B = MRR 32. It is also not an Hα source. Based upon their optical observations, HL85 call HL 8 "simply a reddened star," but the findings at other wavelengths (see below) add complexity. (The only Hα source in the region is HL 11. It is situated on the extreme edge of CCD 7. There were no obvious X-rays at its position, but the sensitivity was low due to vignetting.)

The two X-ray sources are similar in that they are associated with optical nebulosity, IR emission, and continuum radio sources. Their X-ray spectra are typical of the ~$10^7$ K spectra of PMS stars (Table 2) and do not have the soft component characteristic of HH objects. HL 41 at (0.52-1.8) x $10^{31}$ ergs $s^{-1}$ and HL 8 at (4.6-5.0) x $10^{31}$ ergs $s^{-1}$ are high X-ray-luminosity PMS stars assuming they are at 1.7 Kpc. Their continuum radio emissions are also high for PMS or protostars. HL 8 would appear at the high end of both axis and HL 41 would appear off the radio scale of the relationship between quiescent $L_X$ and $L_R$ for all the classes of stars discussed by Güdel (2002, see his Figure 6). The ratios $L_X/L_R$ for HL 41 and 8 are 3.0 x $10^{12}$ and 7.2 x $10^{13}$ Hz, respectively, less than the empirically determined ratio of $10^{14\text{-}16}$ Hz shown by Güdel for a wide variety of stars.

Only the strongest X-ray and radio events, for examples, the giant flares of V773 Tau (Tsuboi et al. 1998) and GMR-A (Bower et al. 2003), have larger X-ray and radio luminosities. There is, however, no evidence that either the radio or X-ray detections of HL 41 or 8 were during flares. Therefore we are left with the interesting result of two apparently quiescent sources that are luminous X-ray and radio sources. MRR 12 (HL 41) has a relatively negative spectral index, -0.8 ± 0.2, which may be indicative of a non-thermal origin. Note finally, that the JHK colors of HL 41 (Figure 9) do not significantly distinguish it from the other sources in the field.

## 5. Conclusions

We report the detection of HH 80 and 81, the brightest X-ray Herbig-Haro objects. Within these objects, there is a strong correspondence at the ≤1700 AU scale between regions of strong X-ray and Hα emission. The luminosities and temperatures revealed in the X-ray spectrophotometry reinforce the model of the large-scale release of energy in the formation of the HH 80/81 complex, although there are no significant X-rays detected at the temperature implied by the faint wings of the optical emission lines. The factor of ~100 higher luminosities compared with the previously detected HH objects bodes well for detection of other powerful galactic outflow sources that are situated favorably with respect to X-ray-absorbing material. A potential relation between X-ray and radio emission needs further exploration.

The region of the central power source of HH 80/81 houses several X-ray sources including a luminous blank-field source near the putative IR-radio power source. Is the X-ray emission a superposition of stellar sources or is it a glimpse of extended emission near the base of the central jet? The region as a whole is clearly an area of star formation based upon the large number of bright X-ray sources detected.

ACKNOWLEDGMENTS

The research described in this paper was performed in part by the Jet Propulsion Laboratory, California Institute of Technology, under contract with the National Aeronautics and Space Administration. We thank Drs. K. Getman and L. Townsley for their assistance in analyzing the ACIS data. We also thank Drs. G. Garmire, L. Rodriguez, B. Reipurth, and E. Feigelson for useful discussions. This research has made use of the NASA/ IPAC Infrared Science Archive, which is operated by the Jet Propulsion Laboratory, California Institute of Technology, under contract with the National Aeronautics and Space Administration. This research has made use of the SIMBAD database, operated at CDS, Strasbourg, France. Some of the data presented in this paper were obtained from the Multimission Archive at the Space Telescope Science Institute (MAST). STScI is operated by the Association of Universities for Research in Astronomy, Inc., under NASA contract NAS5-26555. This publication makes use of data products from the Two Micron All Sky Survey, which is a joint project of the University of Massachusetts and the Infrared Processing and Analysis Center/California Institute of Technology, funded by the National Aeronautics and Space Administration and the National Science Foundation. This research was supported by NASA contract NAS 8-01128.

**Table 1: X-ray Sources (see end)**

**Table 2: Spectral Parameters of Selected X-ray Sources in the HH 80/81 Field**

| CXOPTM Source | Net counts | kT (keV) | $N_H$ ($10^{22}$ cm$^{-2}$) | Abund. (cosmic) | I (< 2 keV) $10^{-15}$erg cm$^{-2}$ s$^{-1}$ | I (> 2 keV) $10^{-14}$ erg cm$^{-2}$ s$^{-1}$ |
|---|---|---|---|---|---|---|
| HH 80/81 composite | 184 | 0.13 ± 0.05 | 0.44 (+ 0.10,-0.12) | > 0.04 | 12 | - |
| J181905.4-205202(HARD) | 30 | > 2.0 | 1.9 (+4.1,-1.3) | 1 (fixed) | 0.93 | 2.5 |
| J181910.4-204657 (HL 41 = MRR12) | 84 | 1.5 (+ 3.0,-0.8) | 1.4 (+0.8,-1.1) | 1 (fixed) | 5.4 | 0.97 |
| J181921.8-204535 (HL 8 = MRR 32) | 223 | >3.0 | 0.15 (+0.45,-0.15 | 1 (fixed) | 29 | 10 |
| J181912.4-204733 (near MRR 14 central source) | 88 | 1.0 (8.0,-0.4) | 20 (±13) | 1 (fixed) | - | 4.7 |
| J181906.2-204232 (0692-0595725 Variable) | 266 | 4.5 (+6.5, -1.7) | 0.24 (+0.17,-0.14) | 1 (fixed) | 19 | 5.3 |
| J181901.8-205242 (SS 365) | 131 | 0.64 (+0.14,-0.16) | 0.68 (+ 0.23, -0.22) | 1 (fixed) | 9.2 | 0.16 |

**Table 3: Physical Properties of HH 80/81 knots**

| | HH 80 | HH 81 | Reference |
|---|---|---|---|
| Angular size(“)[a] | 6 | 4 | 1 |
| Linear size ($10^{17}$ cm) | 1.5 | 1.0 | |
| Volume ($10^{51}$ cm$^3$) | 1.9 | 0.56 | |
| X-ray Temperature ($10^6$K) | 1.5 | 1.5 | |
| X-ray luminosity ($10^{31}$ ergs s$^{-1}$) | 4.5 | 4.3 | |
| Radio flux (mJy)[a] | 1.14 | 1.68 | 1 |
| Hα surface brightness (ergs cm$^{-2}$ s$^{-1}$ arsec$^{-2}$)[b] | 2.5 | 2.1 | 2 |

References.—(1) Martí, Rodríguez, & Reipurth 1993; (2) Heathcote, Reipurth, & Raga 1998

**Table 4: Model Properties of HH 80/81 knots**

| | HH 80 | HH 81 |
|---|---|---|
| Shock velocity (km s$^{-1}$) | 320 | 320 |
| X-ray density (cm$^{-3}$) | 44 | 78 |
| Optical emission preshock density (cm$^{-3}$) | 1100 | 1000 |
| Radio emission preshock density (cm$^{-3}$) | ≤1000 | ≤1000 |

**FIGURE CAPTIONS**

Figure 1. The field in the HH 80/81 complex observed with CCDs S3 and S4. Labels show the regions and some sources discussed in the text.

Figure 2. HH 80 image taken with ACIS-S3. The contours are from the archived HST-WF/PC observation with H-alpha filter taken on 25 August 1995. The red squares are VLA sources from MRR93.

Figure 3. HH 81 image taken with ACIS-S3. The contours are from the archived HST-WF/PC observation with H-alpha filter taken on 25 August 1995. The red squares are VLA sources from MRR93.

Figure 4. Encircled energy vs. distance from centroid for 5 sources detected near the center of ACIS-S3. The two upper plots are X-rays from sources associated with stars; while the three lower plots are from X-ray sources associated with HH knots. The lower axis is units of pixels.

Figure 5. This shows the composite X-ray spectrum for HH 80/81.

Figure 6. CXOPTM J181905.4-205202, hard-spectrum, blank-field X-ray source near HH 80.

Figure 7. The region of GGD 27 and IRS 18162-2048 is shown with the 2MASS $K_s$ image. This area was 5' off-axis on ACIS S3 resulting in the elongation of the X-ray sources shown as green contours with symbols for their centroids. Squares show the location of the VLA sources (MRR93) and larger circles show the locations of IR sources (Stecklum et al. 1997).

Figure 8. The region of GGD 28B is shown with the 2MASS $K_s$ image. The positions of HL 8 (black square, Hartigan & Lada 1985), the VLA source MRR 32 (red square, MRR93), and X-ray source centroid (circle) and contours, are marked.

Figure 9. The *J-H* vs. *H-K* diagram for the X-ray sources detected in the HH 80/81 complex with *Chandra*. The open symbols are the sources further discussed in the text.

Figure 10. The light curve, in 5000 s bins, of CXOPTM J181906.2-204232 (“variable” in the text) = USNO B1 0692-0595725. The X-axis minor tick marks are 2000 s with time measured from the *Chandra* launch.

Figure 11. This shows the relationship between X-ray and radio flux for the four known HH objects with such measurements. (HH 2H from Pravdo et al. 2001, L 1551 from Favata et al. 2002, and HH 80 and 81 from this paper). The function $L_X \sim L_R^{1.4}$ is plotted (solid line).

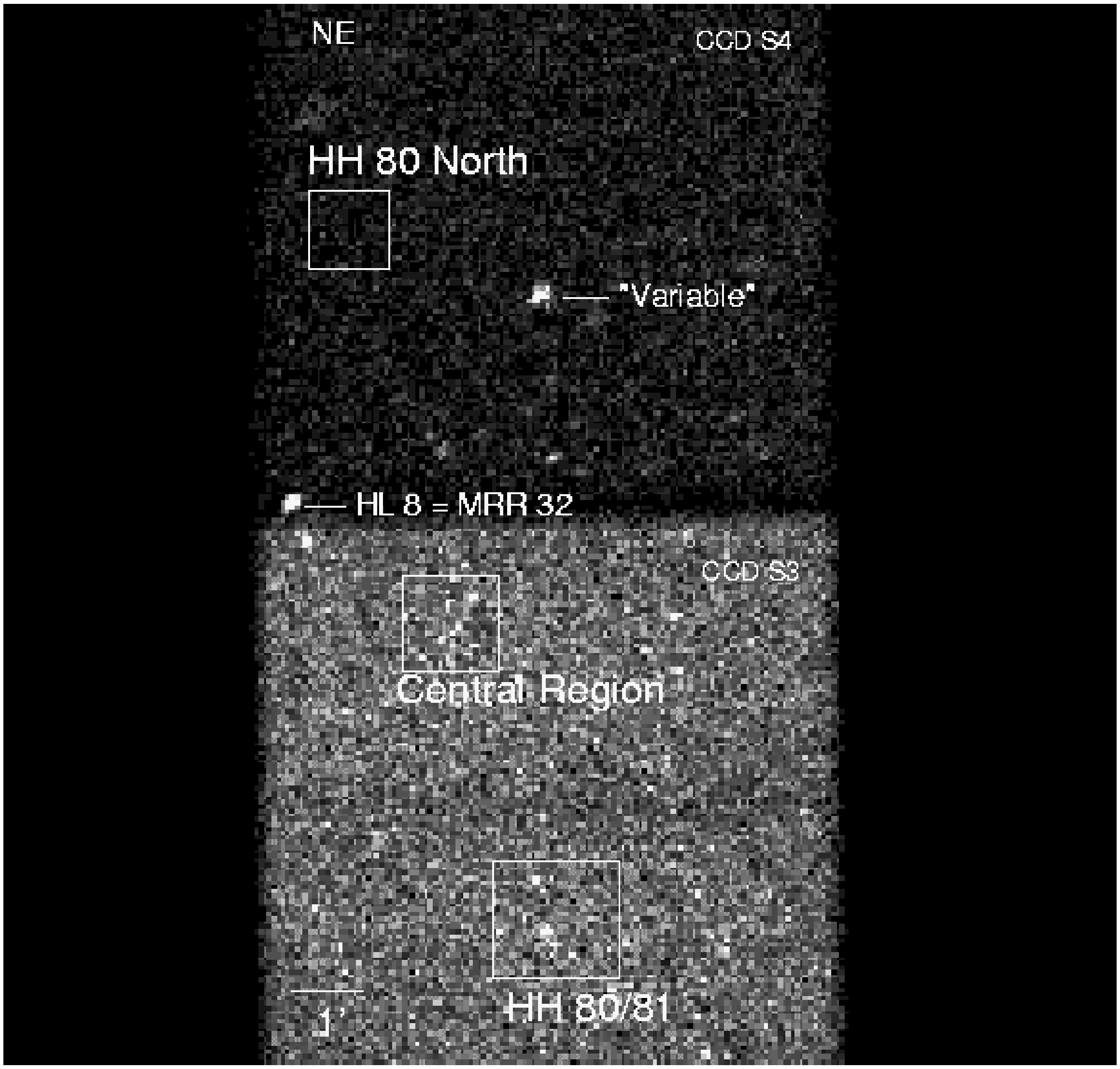

**Figure 1**

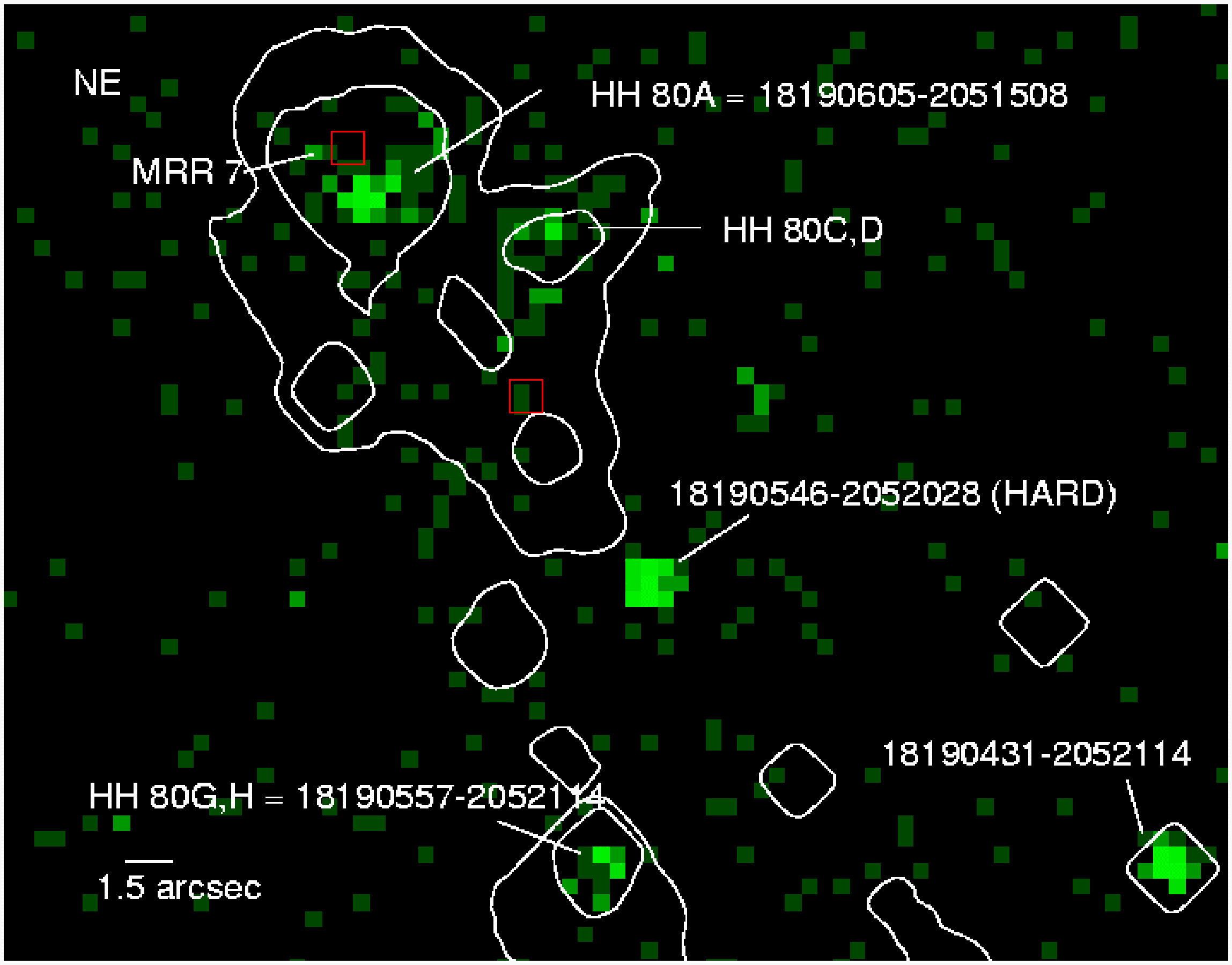
NE
HH 80A = 18190605-2051508
MRR 7
HH 80C,D
18190546-2052028 (HARD)
18190431-2052114
HH 80G,H = 18190557-2052114
1.5 arcsec

**Figure 2**

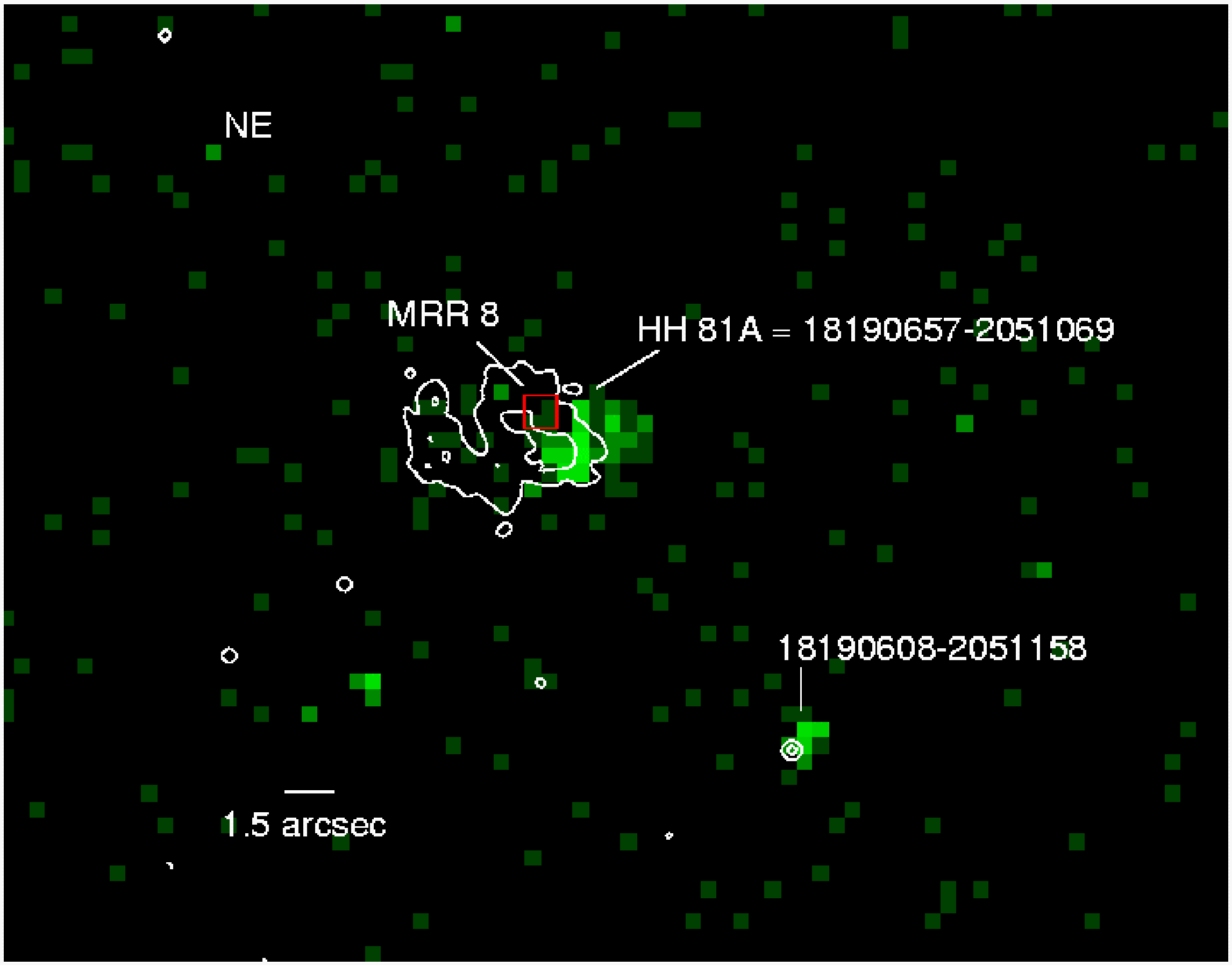

**Figure 3**

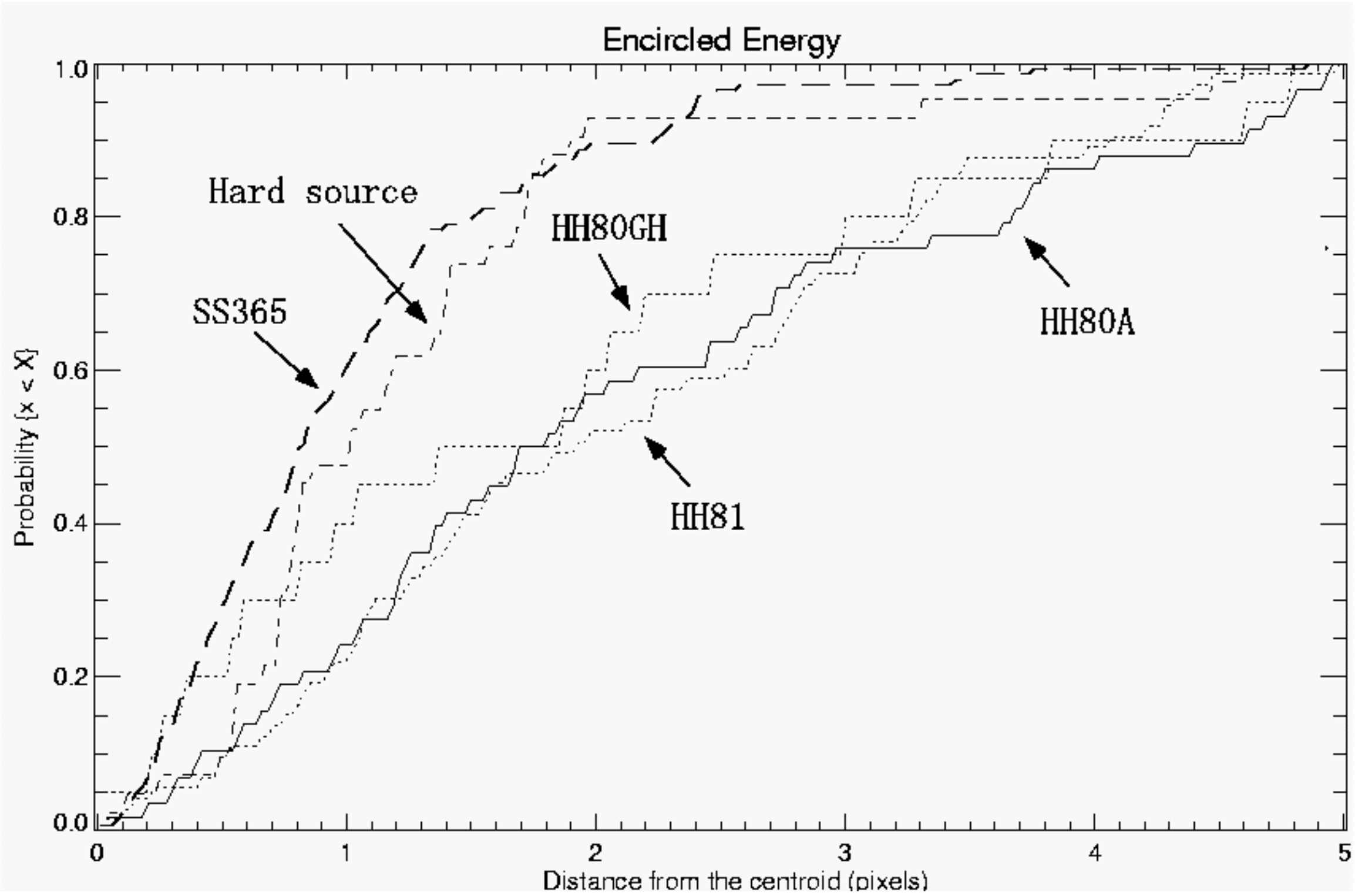

**Figure 4**

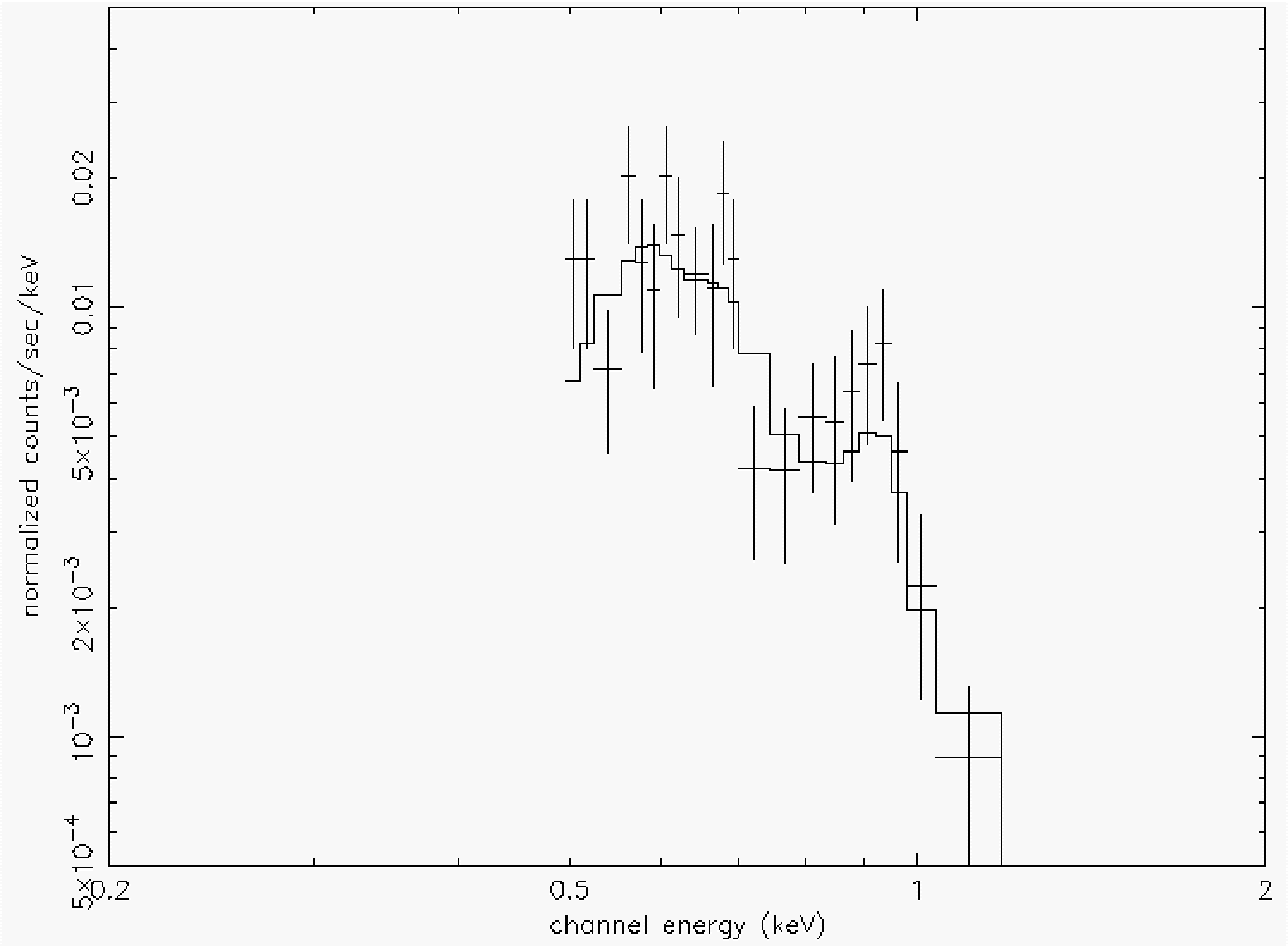

**Figure 5**

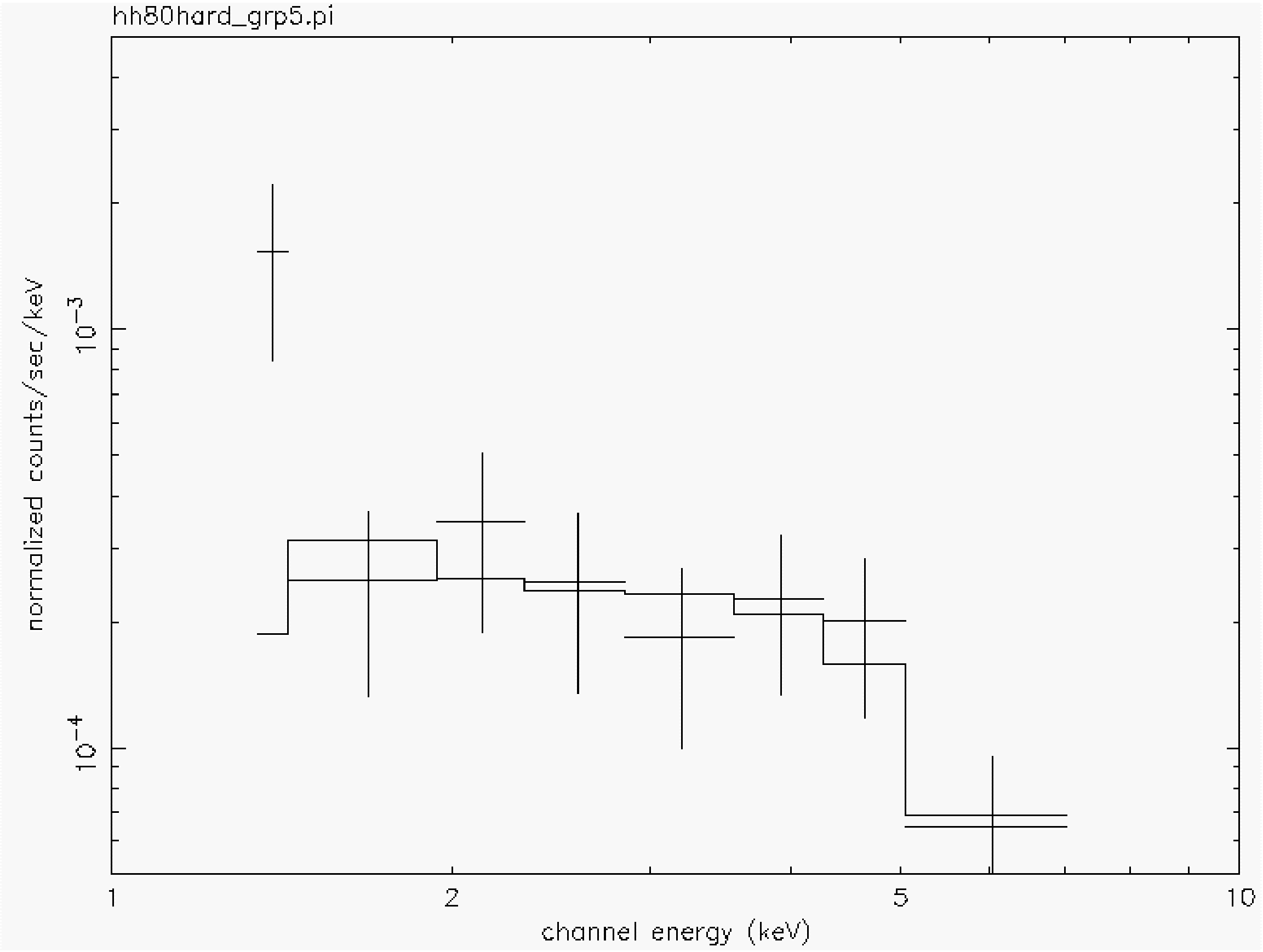

**Figure 6**

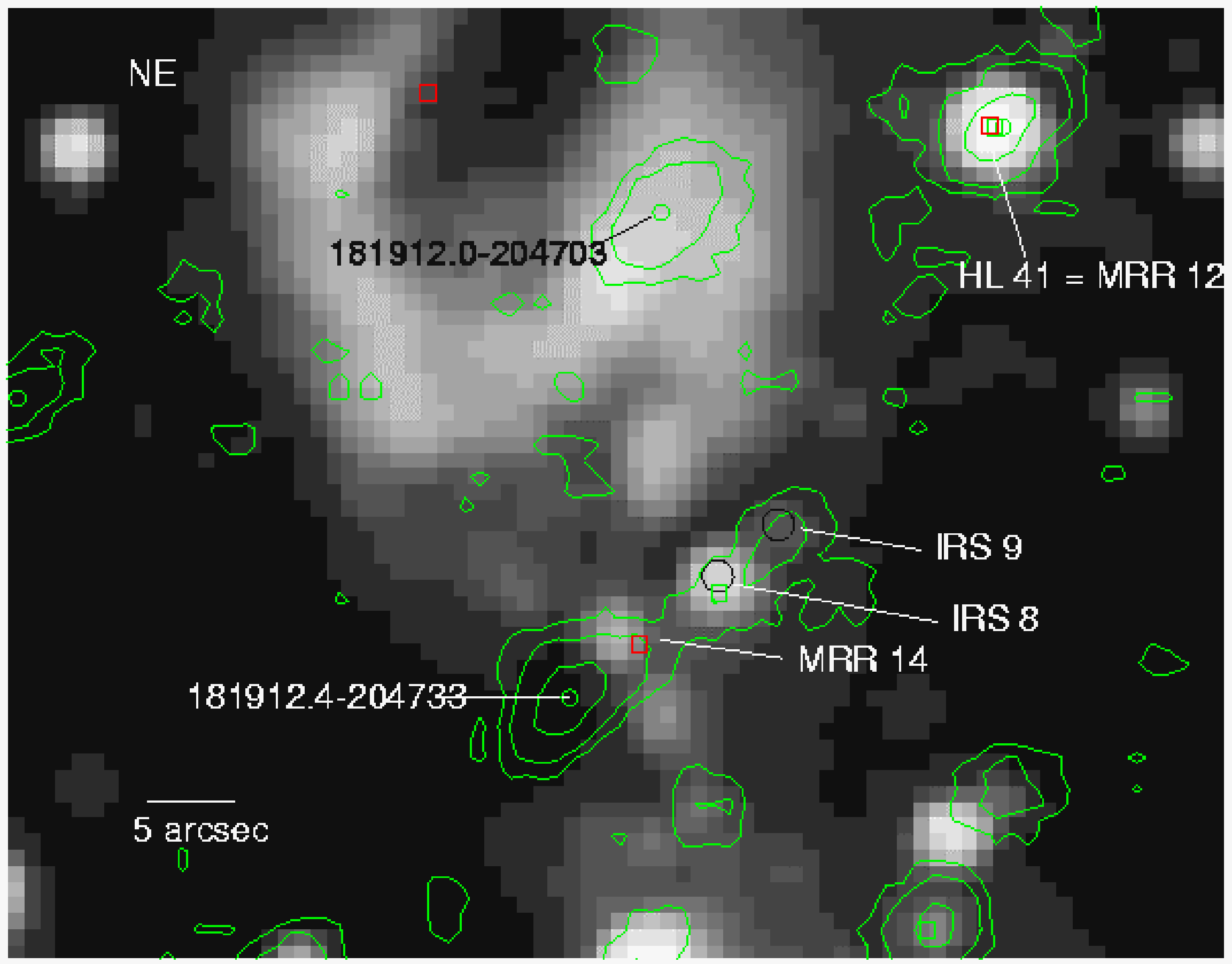

**Figure 7**

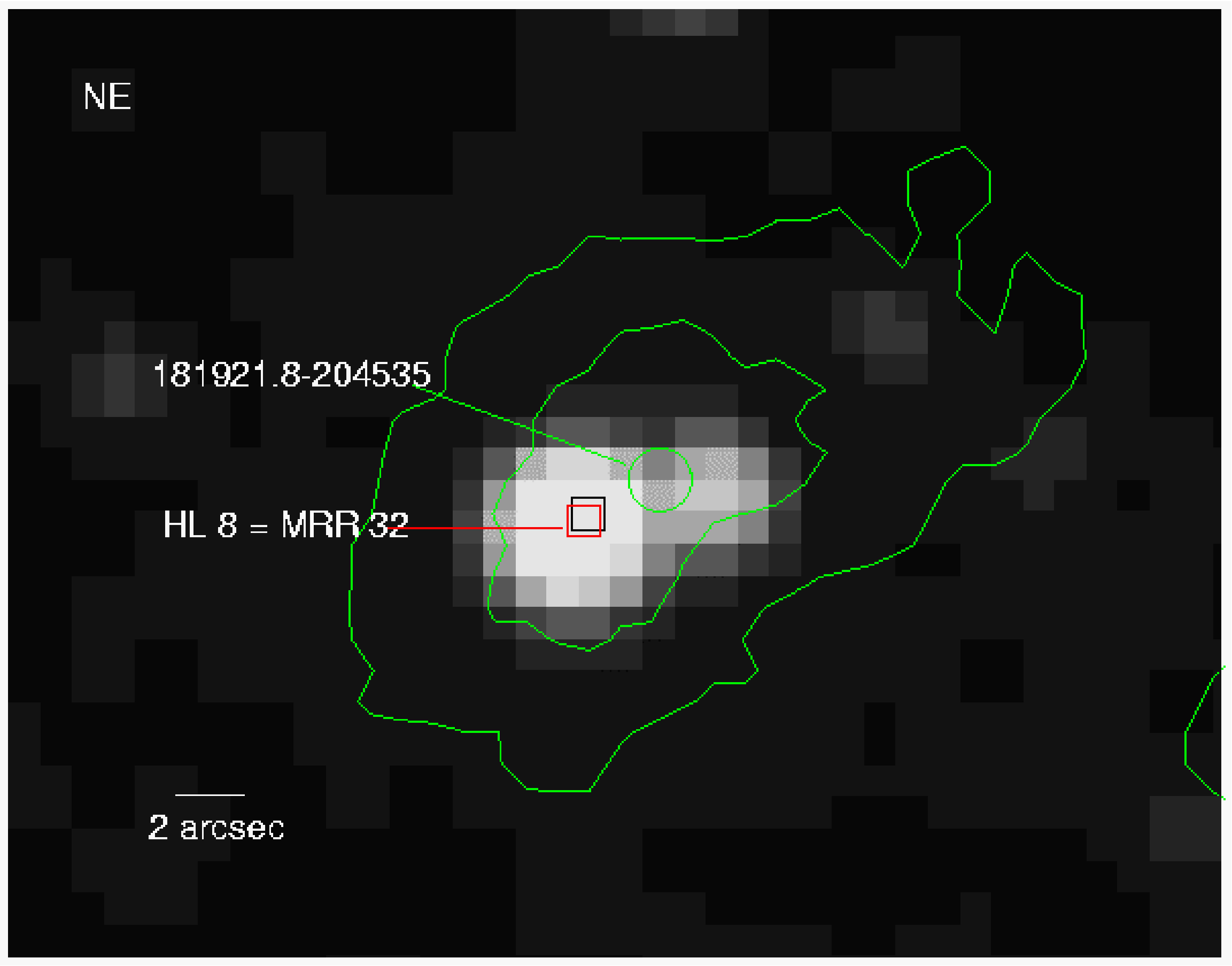

Figure 8

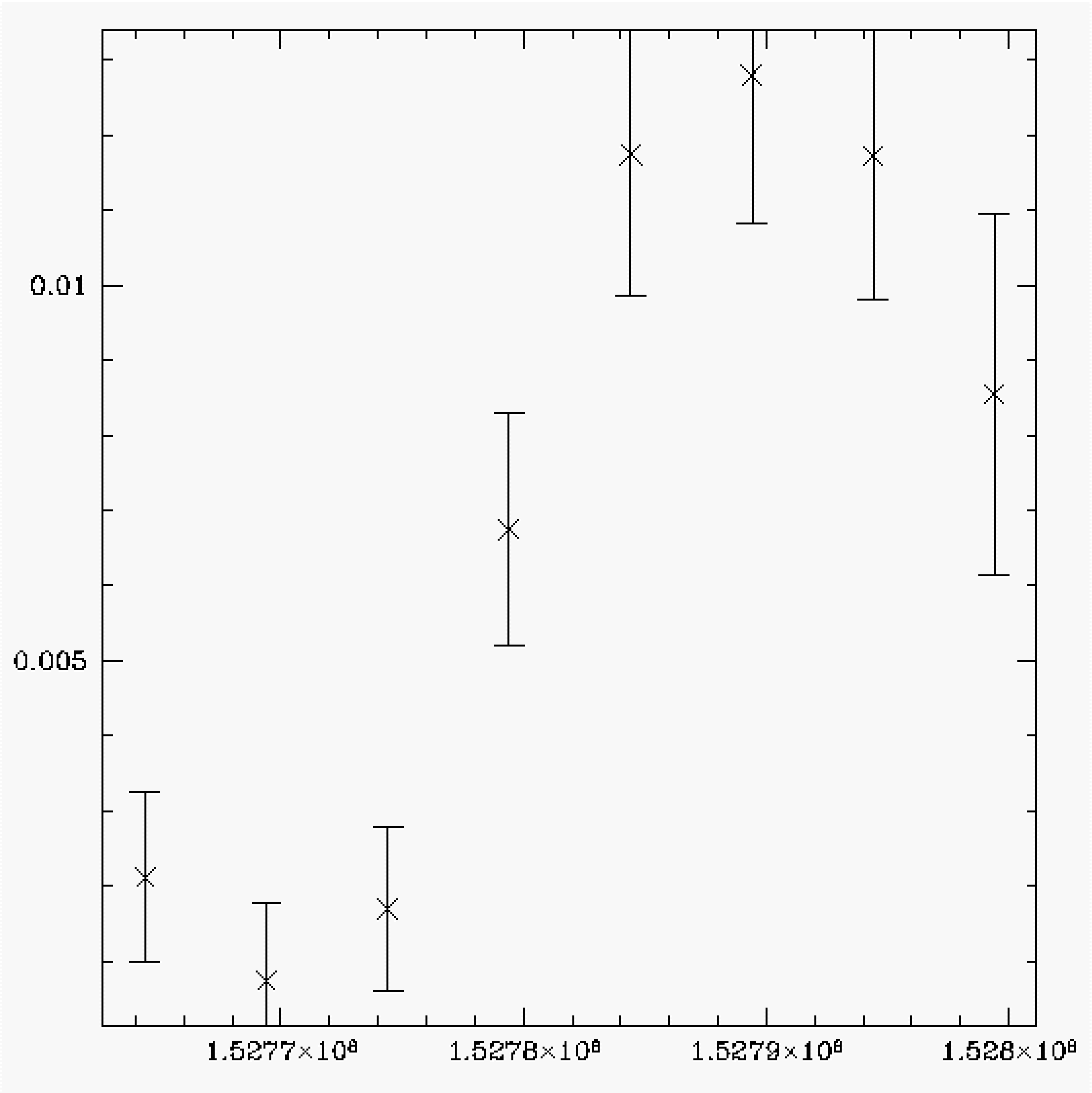

**Figure 9**

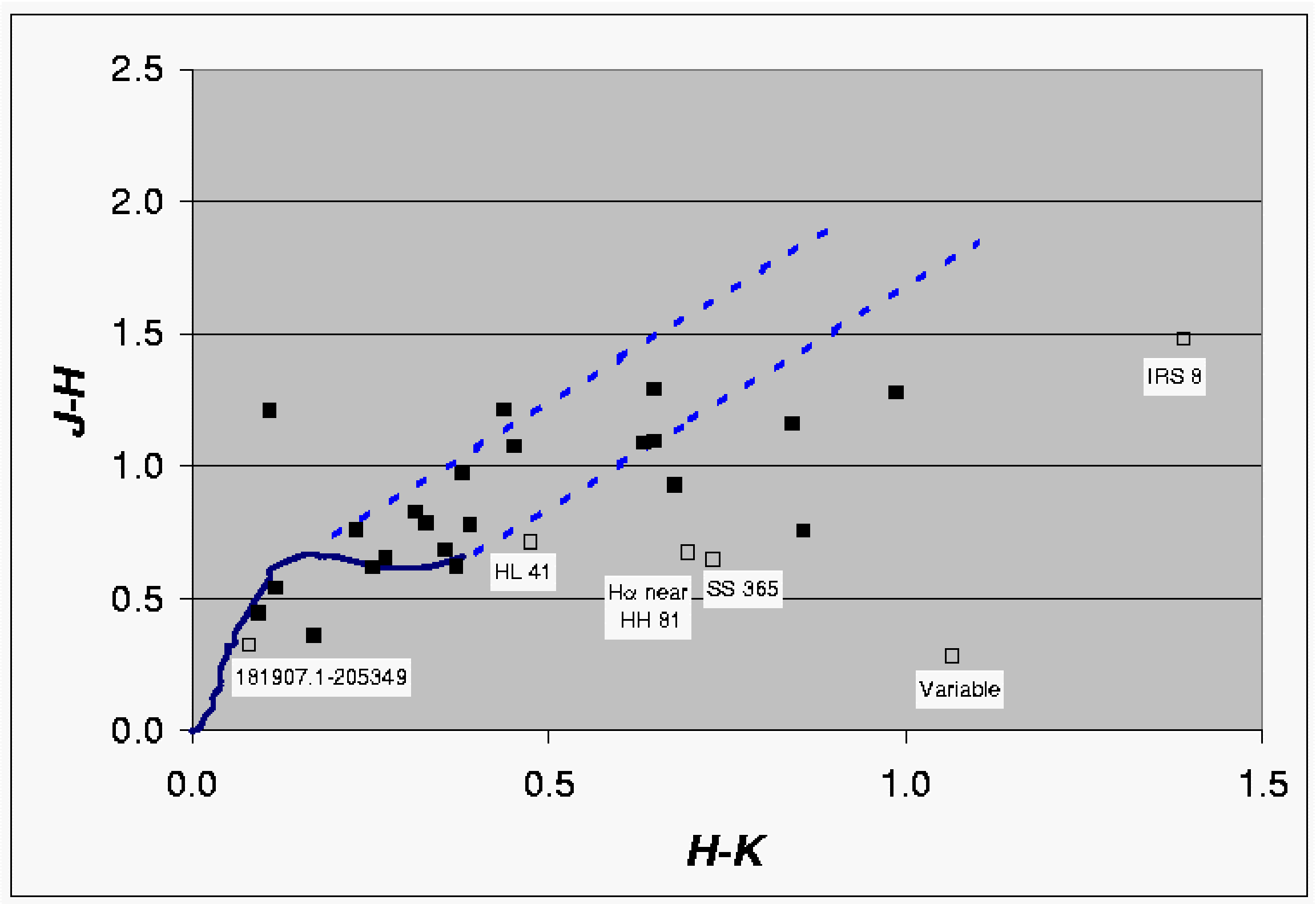

Figure 10

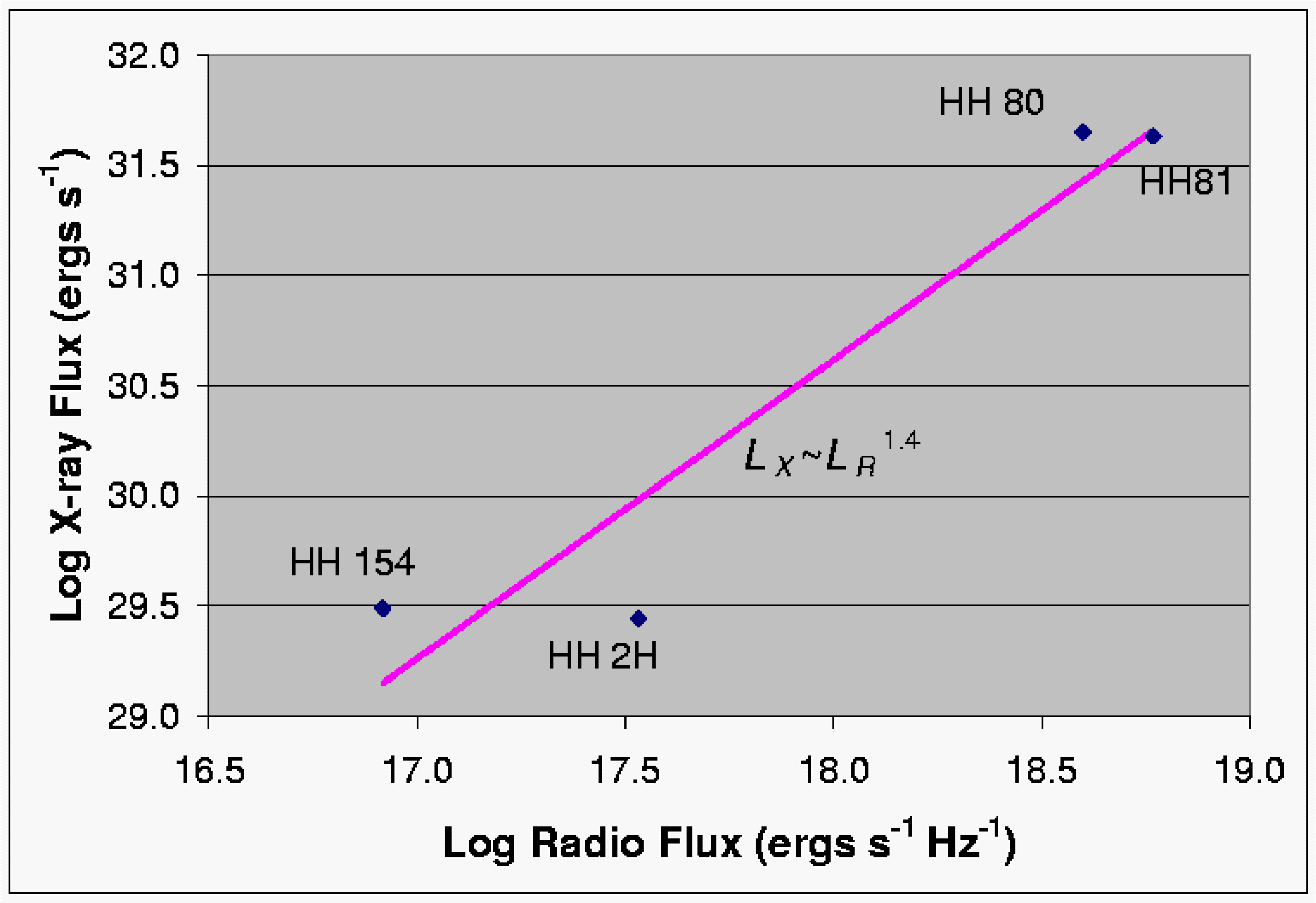

**Figure 11**

**Table 1: X-ray Sources**

| | CXOPTM Name | RA (2000) | Dec.(2000) | Sigma | s c/ks | h c/ks | USNO/2MASS | Sep (") | B | R | I | J | H | K | Note |
|---|---|---|---|---|---|---|---|---|---|---|---|---|---|---|---|
| 1 | J181848.9-205223 | 274.7040 | -20.8733 | 6.2 | 0.38 | | 0691-0603213 | 0.84 | 20.62 | 18.33 | 16.68 | | | | |
| 2 | J181849.1-205137 | 274.7049 | -20.8603 | 3.7 | 0.22 | | | | | | | | | | |
| 3 | J181849.4-204125 | 274.7058 | -20.6905 | 5.9 | 1.31 | f | | | | | | | | | |
| 4 | J181849.4-203947 | 274.7059 | -20.6631 | 3.9 | 0.74 | f | | | | | | | | | |
| 5 | J181849.7-205007 | 274.7075 | -20.8353 | 18.7 | 1.38 | 0.12 | 0691-0603248 | 0.83 | 17.58 | 14.59 | 14.22 | 13.50 | 12.57 | 11.90 | |
| 6 | J181851.5-203735 | 274.7149 | -20.6264 | 5.1 | 1.04 | f | | | | | | | | | |
| 7 | J181852.1-205314 | 274.7172 | -20.8873 | 4.8 | 0.25 | 0.10 | 0691-0603316 | 0.40 | 18.22 | 15.94 | 14.29 | 14.87 | 14.33 | 14.21 | |
| 8 | J181852.3-204453 | 274.7183 | -20.7483 | 4.8 | 0.53 | f | | | | | | | | | |
| 9 | J181853.1-205233 | 274.7215 | -20.8761 | 2.9 | 0.15 | | 0691-0603342 | 0.55 | 18.92 | 16.43 | 15.82 | 14.16 | 13.54 | 13.29 | |
| 10 | J181853.2-205114 | 274.7218 | -20.8540 | 6.6 | 0.38 | | 18185322-2051145 | 0.16 | | | | 15.58 | 14.8 | 14.42 | |
| 11 | J181853.3-205028 | 274.7221 | -20.8413 | 9.4 | 0.55 | | | | | | | | | | |
| 12 | J181853.7-204833 | 274.7241 | -20.8094 | 5.5 | 0.38 | | 0691-0603365 | 1.25 | 20.49 | | 16.66 | 15.23 | 14.54 | 14.19 | |
| 13 | J181854.0-204417 | 274.7254 | -20.7381 | 5.0 | 0.68 | f | | | | | | | | | |
| 14 | J181854.2-205313 | 274.7259 | -20.8871 | 7.7 | | 0.42 | | | | | | | | | |
| 15 | J181854.7-205209 | 274.7281 | -20.8693 | 11.7 | 0.65 | 0.18 | 0691-0603398 | 1.06 | 19.10 | 15.27 | 13.98 | | | | |
| 16 | J181855.1-204658 | 274.7299 | -20.7830 | 3.5 | 0.31 | | | | | | | | | | |
| 17 | J181855.6-205038 | 274.7317 | -20.8442 | 7.9 | 0.44 | | 0691-0603428 | 0.83 | 20.08 | 15.05 | | 15.06 | 13.97 | 13.34 | |
| 18 | J181855.6-205155 | 274.7317 | -20.8653 | 6.2 | 0.31 | 0.15 | 0691-0603433 | 0.47 | 19.99 | 17.61 | 17.49 | 14.45 | 13.5 | 13.15 | |
| 19 | J181855.8-205340 | 274.7328 | -20.8945 | 46.5 | 2.82 | 0.24 | 0691-0603441 | 0.54 | 15.63 | 13.81 | 13.81 | 12.75 | 12.39 | 12.22 | |
| 20 | J181856.1-205235 | 274.7339 | -20.8765 | 6.6 | | 0.36 | | | | | | | | | |
| 21 | J181856.3-205119 | 274.7349 | -20.8553 | 7.2 | 0.39 | 0.13 | 0691-0603456 | 0.64 | | 18.00 | 17.78 | 15.26 | 14.3 | 13.86 | |
| 22 | J181857.1-204603 | 274.7382 | -20.7676 | 12.4 | 1.31 | | 0692-0595582 | 0.72 | 18.63 | 15.96 | 15.62 | 14.14 | 13.48 | 13.21 | |
| 23 | J181857.5-205037 | 274.7396 | -20.8436 | 3.9 | 0.21 | | 0691-0603483 | 0.71 | | 16.65 | 16.44 | 14.62 | 13.8 | 13.55 | |
| 24 | J181857.5-204802 | 274.7398 | -20.8007 | 25.8 | 1.35 | 1.82 | | | | | | | | | |
| 25 | J181857.7-204713 | 274.7407 | -20.7872 | 44.2 | 3.94 | 1.39 | 0692-0595587 | 0.83 | 19.99 | 18.20 | 17.50 | 14.83 | 13.75 | 13.30 | |
| 26 | J181858.1-205006 | 274.7421 | -20.8350 | 2.9 | 0.15 | | 0675-24983619 | 0.41 | 19.60 | 17.30 | | 15.00 | 13.73 | 12.74 | |
| 27 | J181859.2-205230 | 274.7468 | -20.8753 | 2.8 | 0.15 | | 18185924-2052310 | 0.24 | | | | 15.41 | 14.69 | 14.09 | |
| 28 | J181859.4-204946 | 274.7476 | -20.8295 | 10.4 | 0.62 | | 0691-0603559 | 0.31 | 20.31 | 17.00 | 16.00 | 14.76 | 13.97 | 13.64 | |
| 29 | J181859.5-205247 | 274.7482 | -20.8798 | 3.8 | | 0.21 | | | | | | | | | |
| 30 | J181900.7-205241 | 274.7529 | -20.8782 | 7.6 | 0.42 | | 0691-0603595 | 0.42 | 19.90 | 16.65 | | 13.95 | 12.98 | 12.60 | |
| 31 | J181901.0-205202 | 274.7544 | -20.8674 | 3.4 | | 0.18 | | | | | | | | | |

| | CXOPTM Name | RA (2000) | Dec.(2000) | Sigma | s c/ks | h c/ks | USNO/2MASS | Sep (") | B | R | I | J | H | K | Note |
|---|---|---|---|---|---|---|---|---|---|---|---|---|---|---|---|
| 32 | J181901.8-205242 | 274.7575 | -20.8784 | 56.3 | 3.43 | 0.34 | 0691-0603623 | 0.36 | 12.46 | 11.85 | 11.11 | 11.22 | 10.57 | 9.84 | SS 365 |
| 33 | J181903.0-204254 | 274.7626 | -20.7152 | 8.6 | 1.33 | f | | | | | | | | | |
| 34 | J181903.1-204700 | 274.7631 | -20.7836 | 2.7 | 0.18 | | 0692-0595672 | 0.62 | | 17.93 | 17.81 | 15.44 | 13.8 | 13.4 | |
| 35 | J181903.8-204507 | 274.7660 | -20.7520 | 6.3 | 0.62 | f | 0692-0595682 | 0.50 | 19.85 | 18.29 | 16.88 | 14.85 | 13.56 | 12.91 | |
| 36 | J181903.8-204420 | 274.7661 | -20.7391 | 2.9 | 0.25 | f | | | | | | | | | |
| 37 | J181904.3-205211 | 274.7680 | -20.8698 | 20.1 | 1.14 | | 0691-0603673 | 0.27 | 18.80 | 15.85 | 15.75 | 14.52 | 13.77 | 13.54 | |
| 38 | J181904.5-205035 | 274.7688 | -20.8433 | 6.4 | 0.34 | | 18190451-2050358 | 0.04 | | | | 16.22 | 14.70 | 14.49 | |
| 39 | J181905.4-205202 | 274.7728 | -20.8674 | 13.9 | 0.31 | 0.71 | | | | | | | | | HARD |
| 40 | J181905.4-204455 | 274.7729 | -20.7487 | 16.8 | 2.16 | f | 0692-0595712 | 0.72 | 15.44 | 13.63 | 13.54 | 12.17 | 11.73 | 11.63 | |
| 41 | J181905.5-205211 | 274.7732 | -20.8698 | 7.9 | 0.46 | | 0691-0603703 | 0.68 | 18.94 | 16.89 | | | | | HH 80G,H |
| 42 | J181905.5-204745 | 274.7732 | -20.7959 | 2.8 | | 0.19 | | | | | | | | | |
| 43 | J181905.8-204717 | 274.7744 | -20.7881 | 4.2 | 0.31 | | | | | | | | | | |
| 44 | J181905.9-204346 | 274.7747 | -20.7295 | 6.3 | 0.86 | f | | | | | | | | | |
| 45 | J181906.0-205150 | 274.7752 | -20.8641 | 19.8 | 1.23 | | 0691-0603724 | 0.71 | 16.68 | 14.10 | 17.46 | | | | HH 80A |
| 46 | J181906.0-205115 | 274.7753 | -20.8544 | 7.9 | 0.44 | | 0691-0603726 | 0.51 | 18.98 | 16.01 | 16.08 | 15.13 | 14.46 | 13.76 | HH 81 knot? |
| 47 | J181906.1-205011 | 274.7757 | -20.8366 | 9.6 | | 0.53 | | | | | | | | | |
| 48 | J181906.1-205338 | 274.7757 | -20.8940 | 14.9 | 0.37 | 0.74 | | | | | | | | | |
| 49 | J181906.2-204232 | 274.7761 | -20.7090 | 30.5 | 7.26 | f | 0692-0595725 | 0.36 | 19.38 | 15.48 | | 12.74 | 12.46 | 11.40 | Variable |
| 50 | J181906.2-204919 | 274.7762 | -20.8221 | 4.1 | 0.23 | | 0691-0603730 | 0.71 | 19.99 | 17.55 | 17.04 | 14.85 | 13.76 | 13.11 | |
| 51 | J181906.5-205106 | 274.7774 | -20.8519 | 26.4 | 1.68 | | | | | | | | | | HH 81A |
| 52 | J181907.0-205114 | 274.7794 | -20.8540 | 2.5 | | 0.13 | | | | | | | | | |
| 53 | J181907.1-205349 | 274.7798 | -20.8972 | 4.3 | 0.23 | | 0691-0603750 | 1.41 | 14.62 | 13.06 | 12.84 | 13.12 | 12.79 | 12.71 | |
| 54 | J181907.2-205138 | 274.7801 | -20.8608 | 6.6 | 0.37 | | 0691-0603751 | 0.03 | | 17.87 | 17.62 | 15.33 | 14.36 | 13.30 | |
| 55 | J181907.3-205356 | 274.7808 | -20.8990 | 10.6 | 0.58 | | | | | | | | | | |
| 56 | J181907.6-204809 | 274.7819 | -20.8027 | 34.1 | 2.63 | 0.25 | 0691-0603761 | 0.31 | 19.10 | 15.75 | 15.26 | 13.30 | 12.52 | 12.13 | |
| 57 | J181908.4-204039 | 274.7854 | -20.6777 | 3.9 | 0.84 | f | | | | | | | | | |
| 58 | J181908.8-205151 | 274.7867 | -20.8644 | 10.3 | | 0.56 | | | | | | | | | |
| 59 | J181909.1-204716 | 274.7883 | -20.7879 | 4.3 | 0.31 | | | | | | | | | | |
| 60 | J181909.6-204832 | 274.7902 | -20.8089 | 2.8 | 0.17 | | 18190968-2048329 | 1.05 | | | | 16.68 | 15.47 | 15.36 | |
| 61 | J181909.8-204431 | 274.7909 | -20.7420 | 4.4 | 0.44 | f | | | | | | | | | |
| 62 | J181910.4-204657 | 274.7937 | -20.7828 | 16.3 | 1.45 | 0.70 | 0692-0595779 | 0.81 | 12.61 | 11.74 | 10.54 | 11.21 | 10.50 | 10.02 | MRR 12 |
| 63 | J181910.8-204748 | 274.7951 | -20.7968 | 12.5 | 1.05 | | 18191078-2047481 | 0.59 | | | | 14.97 | 13.84 | 12.85 | |
| 64 | J181910.9-204823 | 274.7955 | -20.8065 | 2.0 | | 0.12 | | | | | | | | | |

| | CXOPTM Name | RA (2000) | Dec.(2000) | Sigma | s c/ks | h c/ks | USNO/2MASS | Sep (") | B | R | I | J | H | K | Note |
|---|---|---|---|---|---|---|---|---|---|---|---|---|---|---|---|
| 65 | J181911.2-204631 | 274.7970 | -20.7755 | 2.9 | | 0.22 | | | | | | | | | |
| 66 | J181911.7-204727 | 274.7990 | -20.7909 | 3.0 | 0.21 | | 18191175-2047264 | 0.77 | | | | 13.95 | 12.47 | 11.08 | IRS 8 |
| 67 | J181911.7-204803 | 274.7991 | -20.8010 | 3.3 | 0.24 | | | | | | | | | | HL 31? |
| 68 | J181911.8-204740 | 274.7994 | -20.7946 | 3.6 | | 0.26 | 18191184-2047406 | 0.14 | | | | 15.9 | 14.1 | 11.12 | |
| 69 | J181912.0-204703 | 274.8001 | -20.7842 | 13.6 | | 1.33 | | | | | | | | | |
| 70 | J181912.0-205330 | 274.8001 | -20.8919 | 4.9 | 0.26 | | 0691-0603827 | 1.26 | 15.86 | 14.72 | 13.80 | 13.38 | 12.69 | 12.34 | |
| 71 | J181912.4-204445 | 274.8018 | -20.7460 | 9.6 | 1.07 | f | | | | | | | | | |
| 72 | J181912.4-204733 | 274.8018 | -20.7927 | 21.7 | | 2.30 | | | | | | | | | 3.8"MRR 14 |
| 73 | J181912.9-205253 | 274.8039 | -20.8816 | 2.5 | 0.13 | | | | | | | | | | |
| 74 | J181913.2-204436 | 274.8052 | -20.7436 | 6.0 | 0.58 | f | 0675-25004667 | 1.27 | 19.50 | 17.60 | | 14.42 | 13.80 | 13.43 | |
| 75 | J181913.4-205116 | 274.8061 | -20.8546 | 9.6 | 0.55 | | 18191348-2051166 | 0.33 | | | | 15.40 | 13.94 | 12.96 | |
| 76 | J181913.6-204750 | 274.8069 | -20.7974 | 4.9 | 0.33 | 0.13 | 18191364-2047505 | 0.21 | | | | 14.78 | 13.09 | 12.06 | |
| 77 | J181913.6-204751 | 274.8071 | -20.7975 | 4.4 | 0.23 | | 18191364-2047505 | 0.98 | | | | 14.78 | 13.09 | 12.06 | |
| 78 | J181913.9-205331 | 274.8081 | -20.8922 | 4.3 | | 0.26 | 0691-0603839 | 1.10 | 21.43 | 17.49 | 18.17 | 15.69 | 14.93 | 14.07 | |
| 79 | J181914.0-205227 | 274.8084 | -20.8744 | 4.8 | | 0.23 | | | | | | | | | |
| 80 | J181914.3-204947 | 274.8098 | -20.8298 | 4.2 | | 0.52 | | | | | | | | | |
| 81 | J181914.9-204714 | 274.8121 | -20.7875 | 6.4 | 0.24 | | 18191483-2047144 | 1.10 | | | | 17.75 | 16 | 14.74 | |
| 82 | J181915.0-204908 | 274.8125 | -20.8189 | 4.0 | 0.13 | 0.21 | | | | | | | | | |
| 83 | J181916.7-205025 | 274.8197 | -20.8404 | 3.3 | 0.18 | | 18191668-2050247 | 0.90 | | | | 15.77 | 14.1 | 13.5 | |
| 84 | J181918.7-205205 | 274.8283 | -20.8683 | 5.8 | 0.21 | 0.23 | | | | | | | | | |
| 85 | J181919.5-205220 | 274.8316 | -20.8725 | 2.7 | 0.15 | | | | | | | | | | |
| 86 | J181920.1-203951 | 274.8339 | -20.6642 | 10.6 | 2.88 | f | | | | | | | | | |
| 87 | J181920.4-205131 | 274.8351 | -20.8588 | 31.7 | 2.08 | 0.15 | 0691-0603896 | 0.95 | 19.76 | 15.89 | 15.45 | 13.32 | 12.50 | 12.18 | |
| 88 | J181920.7-204633 | 274.8365 | -20.7761 | 3.1 | | 0.23 | | | | | | | | | |
| 89 | J181920.9-204609 | 274.8374 | -20.7694 | 20.1 | 2.35 | 0.76 | 0692-0595859 | 0.42 | 20.41 | 16.43 | 15.89 | 13.08 | 11.92 | 11.08 | |
| 90 | J181921.2-205054 | 274.8386 | -20.8485 | 2.6 | | 0.15 | | | | | | | | | |
| 91 | J181921.8-204535 | 274.8412 | -20.7598 | 46.0 | 6.57 | f | Confused in 2MASS | | | | | | | | 2.7"MRR 32 |
| 92 | J181922.8-204645 | 274.8452 | -20.7792 | 6.0 | 0.48 | | 18192276-2046451 | 1.15 | | | | 14.97 | 14 | 13.5 | |
| 93 | J181923.3-204147 | 274.8474 | -20.6966 | 2.8 | 0.30 | f | 18192336-2041477 | 0.20 | | | | 15.52 | 14.85 | 14.03 | |
| 94 | J181923.4-205210 | 274.8479 | -20.8694 | 2.0 | 0.07 | 0.08 | 0691-0603921 | 0.19 | 20.69 | 17.69 | 18.46 | 15.71 | 14.49 | 14.06 | |